\documentclass[journal]{IEEEtran}
\usepackage{cite}
\usepackage{amsmath}
\usepackage{amsfonts}
\usepackage{array}
\usepackage{multirow}
\usepackage{amssymb}
\usepackage{graphicx}
\usepackage{multirow}
\usepackage{amsthm}
\usepackage{amsbsy}

\usepackage{epstopdf}
\newtheorem{remark}{$\mathbf{Remark}$}
\title{A Decentralised Control Strategy for Secondary Voltage Regulation}
\author{Jitendra Kumar Goyal, Vinu Thomas, Bogdan Marinescu
\thanks{The authors are with Ecole Centrale de Nantes, France--44300 (email: Jitendra-kumar.Goyal@ec-nantes.fr, Vinu.Thomas@ec-nantes.fr, Bogdan.Marinescu@ec-nantes.fr)}
}
\begin{document}
\maketitle
\begin{abstract}
This paper proposes a decentralised secondary voltage control strategy that has several benefits over the existing centralised strategies. For that, a new structure for the control is proposed in terms of an inner and outer loops for each generator. The individual generators of a particular zone  participate in the secondary voltage control by aligning their reactive powers with respect to the pilot point voltage reference. The decentralised nature of the proposed control strategy enables plug and play operation: run with different numbers of generators without any need of regulators reconfiguration, resilience in case of generator failure. This allows one to use such control to integrate renewable generators to existing secondary regulations alongside with the classic generators. The proposed control strategy is implemented using a model-free control scheme that does not require a higher order complex model of the power grid. The deployment of a discrete-time intelligent proportional controller simplifies the tuning of the controller gains. The proposed strategy is validated on a four generator power system model  in MATLAB/Simulink/Simelectrical environment. Simulation results are presented to show the effectiveness of the proposed strategy along with different case studies such as load perturbation, transmission line perturbation, generator  disconnection and delay in pilot point voltage measurement to highlight its robustness. 
\end{abstract}

\section{Introduction}
Modern power grids consist of large number of generators and loads interconnected by a complex network of transmission and distribution lines. It is essential to ensure that the voltage and frequency of the grid are maintained within its operating range to have a stable and reliable operation of the power grid. Usually, a hierarchical control system, consisting of three levels (primary, secondary and tertiary), is used to regulate the voltage levels at various nodes of the grid\cite{taranto,b4,b1}. In the primary level of voltage control, automatic voltage regulators (AVRs) are deployed at the generating stations to ensure the fast regulation of the terminal voltages of the synchronous generators, by controlling the field excitation. The typical time constant for this control loop is less than a second. Whenever the voltage levels at different nodes or buses in the grid deviates from its nominal values, a secondary voltage control (SVC) operates by providing a correction signal to the AVRs.
For the purpose of secondary voltage control, the power system network is divided into different zones and the voltage regulation in each zone is implemented by controlling the reactive power flow of the generators participating in the secondary voltage control of that zone. The dynamics of the secondary voltage control (around 2 to 3 minutes)  is slower than the AVR dynamics. A pilot point bus is chosen for each zone, such that the voltage of the pilot point gives an indication of the voltage levels prevailing in that zone.  The SVC maintains the voltage level of the pilot point in each zone at the pilot point reference voltage \cite{b25}.  The tertiary voltage controller determines the optimal voltage profile of the network and provides the reference pilot voltage reference for the secondary voltage control. The optimal voltage profile is decided according to the prevailing supply demand conditions to meet the economic and safety requirements.
\par In the classical SVC \cite{taranto}, a proportional-integral(PI) controller for each zone generates a signal proportional to the required reactive power level to maintain the pilot point voltage to the reference level. The local reactive power controllers of the  generators operate on the basis of the PI controller output signal and the participation factors of the individual generators. The reactive power controller provides a correction signal to the AVR of the generators. The decoupling of the time scales of AVR and the reactive power control loops is essential to avoid the interactions between the loops, but it also leads to a non-minimum phase behaviour that may lead to transient stability \cite{b4} . The coordinated automatic control of reactive powers from the generators in a zone  \cite{b1} helps to address this issue and is effective in managing the voltages in larger zones of the grid. An inter-plant voltage control scheme was proposed in \cite{b26} for the optimised voltage control of a series of power plants by co-ordinating their reactive power reserves.
In \cite{b3}, a new strategy for SVC was proposed that is easier to implement and has robust performance. But, this SVC strategy being  fully model based, large changes in the system parameters would adversely affect the controller performance. Also, some of the above-mentioned existing strategies involve the optimization of a multi-variable quadratic function \cite{b23}. Moreover, when the tracking objectives for voltage and reactive power are mixed into a quadratic function that is to be minimised, it results in uncertain reactive power alignment. \par Most of the existing SVC strategies are centralised and they operate satisfactorily when the real-time measurements are received by the controller from the individual generators. A communication delay or even a failure or non-participation of an individual generator can adversely affect the stability of the centralised SVC strategies. A decentralised structure for the SVC would enable the individual generators to participate in the SVC with minimum external information. The reactive power alignment of the participating generators based on the local information would make the SVC scheme resilient against uncertainties. A decentralised type of control which operates on the individual generators to participate in the SVC and manages to maintain the pilot point voltage at the required level would be highly beneficial. The benefits of a decentralised SVC control scheme are its resiliency against generator disconnection and communication delays or failures. 
\par Motivated by aforementioned observations, this work presents a newly decentralised control strategy for the secondary voltage control that has potential applications in power grids with multiple zones and with higher participation of renewable energy generators in the secondary voltage control. In contrast to the existing centralised SVC control schemes, the proposed control strategy operates in a decentralised manner at the individual generators participating in the SVC by acquiring the information about the reference and actual pilot point voltages. Thereafter, the proposed scheme enables the generators to align their reactive power contributions with respect to the reference pilot point voltage. This type of control is highly beneficial when there are some unforeseen events such as a generator disconnection and communication failure between the individual generators and the SVC controller. A benchmark system consisting of four generators is considered and different case studies such as load perturbation, transmission line perturbation, generators disconnection, communication delays, etc are conducted on the benchmark system to show the efficacy and robustness of the proposed SVC scheme.
\par The paper is structured as follows. Section II gives a brief overview of the classic SVC and the need for an improvement. The use of SVC contribution block, concepts of the newly proposed SVC strategy and controller synthesis are discussed in detail in section III. The benefits of the proposed SVC method is also presented. Section IV presents the details of a test system to verify the proposed SVC strategy. The MATLAB/Simulink results obtained for various cases are illustrated and discussed. The conclusions ad future directions of the proposed control scheme are given in Section V. 
\section{Problem Statement and Preliminaries}
The secondary voltage control dynamics are much slower than that of primary voltage control (AVR) control dynamics. Therefore, neglecting the local electric dynamics, the power system dynamics under the SVC action can be analyzed and modeled with the help of sensitivities matrices, $C_v$ and $C_q$. $C_v\in \mathbb{R}^{1 \times n}$ and $C_q \in \mathbb{R}^{n \times n}$ represent the sensitivities matrices of the pilot point voltage $V_{pp}$ and reactive powers $Q=\begin{bmatrix} Q_1, & \cdots, & Q_n \end{bmatrix}^T$ with respect to the terminal voltage of the generators $V_{t}=\begin{bmatrix} V_{t_1}, & \cdots, & V_{t_n} \end{bmatrix}^T$, respectively, where $n$ denotes the number of generators that participate in the secondary voltage control. Pilot point voltage $V_{pp}$ is assumed to be a scalar quantity since zones with only one pilot point voltage is considered throughout this work, although it can be extended to grids with multiple zones.  \par Based on the above discussion, in general, the power system dynamics under SVC action is modeled as:
\begin{align}\label{s1}
	V_{pp}&=C_vV_{t} \\\label{s2} Q&=C_qV_{t}
\end{align} 
\subsection{SVC control objectives}
Two control objectives of SVC are
\begin{enumerate}
	\item Set-point tracking of pilot point voltage, i.e.,
	\begin{equation}\label{ob1}
		V_{pp}(t) \rightarrow V_{pp}^{ref}, ~\text{as} ~t \rightarrow \infty
	\end{equation}
	\item Managing (align) the reactive power of each generator that participate in SVC action.
\end{enumerate}
The objective of the proposed SVC strategy is to ensure the tracking of the pilot point reference voltage by aligning the reactive powers of the SVC participating generators. 

Before proceeding to the main contribution of this work, we will revisit the classic SVC, its challenges and limitations.
\subsection{Classic SVC and its limitations}
\begin{figure}[!h]
	\centering
	\includegraphics[width=8cm]{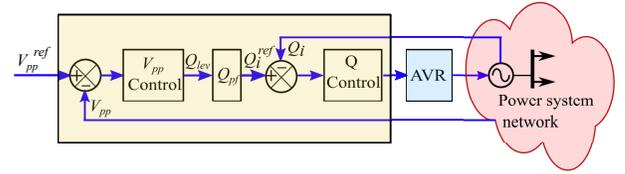}
	\caption{Block diagram of classic SVC strategy}
	\label{classic}
\end{figure}
The block diagram of classic SVC strategy is shown in Fig. \ref{classic}. The working principle of classic SVC is detailed in \cite{b3}.  It has also been discussed in the introduction section. In this section, a brief overview of classic SVC is presented to highlight its challenges and limitations. In classic SVC, a signal denoting the required reactive power level from each zone is computed which is further applied to adjust the set point of primary voltage loop of AVR's with the help of reactive control loop that regulates each $Q_i(t)$ of the generators for the alignment of its steady-state value with other generating units as per the participation factor $Q_i^{pf}$, indicated for each generators. For smooth operation, better implementation and to avoid interactions, the local reactive power loop and voltage control loop are time-decoupled, i.e.  local reactive power control loop is faster than the voltage control loop. However, classic SVC suffers from non-minimum phase response due to its hierarchical structure. It may lead to transient instability which is clearly explained in \cite{b4}. \par To overcome this drawback, coordinated SVC has been introduced in \cite{b12,b13} where the voltage stability margin has been improved by coordinating reactive power reserves through introduction of inter-plant voltage control. However, such techniques have faced a lot of challenges in real-time implementation due to its dependence on online-optimization of multi-variable quadratic function \cite{b23}. One of the major drawbacks of this technique is that reactive power alignment is not possible since reactive and voltage tracking objectives are mixed together to form the quadratic function which is to be minimized.\par Further, another SVC strategy was proposed in \cite{b3} to overcome the issues and limitations occurring in classic SVC and coordinated SVC. A SVC solution has been provided to take over the problem of pilot point voltage control and alignment of reactive power of each generating units by means of average tracking of reactive power rather than the constant reference value. A multi-variable proportional integral controller has been used to achieve the SVC objectives. It incorporated LQR method \cite{b15} for the computation of the PI controller gains. However, to the best of authors knowledge, till now no such methods were available in the literature that could be used easily and effectively,in the computation of multi-variable PI controller gains . Also, the control structures implemented above are centralised which limits the applications. 
\par The aforementioned discussion shows that a lot of improvement is still required and need to be carried out in the area of secondary voltage control. Taking this into consideration, in this work, a new SVC solution is proposed which is fully decentralised with the objective to track the reactive power of each generators and regulate the pilot point voltage under uncertainties and external disturbances such as load perturbation, transmission line perturbation, generators reconnection and disconnection, and large delay margin in pilot point voltage to highlight its robustness feature. The proposed approach take over the advantage of all the existing SVC's strategy as discussed above (classic SVC, coordinated SVC and dynamically feasible SVC). The proposed SVC strategy includes the design of a SVC contribution block which is used for generation of signal $Q_i^{ref}$ (defined later) for the alignment of reactive power and implementation of model-free control scheme with decentralized Discrete-Time intelligent Proportional (DTiP) controller to achieve the SVC objectives. In this work, DTiP controller overcome the drawback of multi-variable PI controller as discussed earlier. Model-free control strategy is adopted in the proposed SVC solution due to its inherent properties like simplicity, simplified parametrization, fast and easy implementation and great robustness with respect to numerous external disturbances. One of the important feature of our proposed SVC scheme is that it can tolerate and show robustness against a large value of delay margin in the pilot point voltage and the control structure is completely decentralised whereas the SVC strategy with multi-variable PI controller fails to do so. This fact is clearly discussed and demonstrated in section IV.

\section{Main Contributions}
This section presents the main contributions of this work. A new SVC strategy has been proposed here to overcome the limitations of the existing SVC strategies. While the existing strategies implement a centralized SVC strategy, the proposed method is a decentralised SVC strategy, that operates based on the information from the pilot point and the local measurements only. The real-time measurements from the other generators participating in the SVC is not necessary for its operation. The outer loop of the proposed control strategy generates a control signal, based on the information of $V_{pp}(t)$ and $V_{pp}^{ref}$ and it is added to the  output of SVC contribution block to generate a reactive power reference signal $Q_i^{ref}$ for the individual generators. The inner reactive power control loop then operates to regulate the reactive power of the individual generators such that $V_{pp}(t)$ is maintained at $V_{pp}^{ref}$. \par Before proceeding to the new SVC proposed strategy in detail, we will first discuss about the SVC contribution block since it plays an important role in achieving the SVC objectives in the newly proposed strategy.
\subsection{SVC contribution}
This block is used for the generation of signal, $Q_i^{ref}$ for each generators from the information of $V_{pp}^{ref}$. The generated signal, $Q_i^{ref}$ is further used for the computation of $Q_{i}^{ref}$ for the inner loop with the aim of regulating the $Q_i$ in order to align its steady-state value with the others generator units, as per the participation factors $Q_i^{pf}$ indicated for each units. It is clearly shown in Fig. \ref{outer}. In this work, reactive power alignment refers to reactive power control of all the generators which participate in the SVC under the steady state conditions, i.e., after vanishing of all the transients from the response and reach the steady state. The computation of signal, $Q_i^{ref}$ from $V_{pp}^{ref}$ is described as follows.
\par It is well known \cite{b3} at steady state, the reactive power alignment satisfies the following condition.
\begin{equation}\label{f1}
\frac{Q_i}{Q_i^{pf}}=c
\end{equation} where $c$ is any constant.
Based on \eqref{f1}, one can write $Q_i=Q_i^{ref}$ where the signal $Q_i^{ref}$ represents the steady state value of reactive power of each generators which are participating in SVC action .  Then, from the relation \eqref{s1} and \eqref{f1}, the following equations can be rewritten.
\begin{equation}\label{f2}
\Sigma=\begin{cases}
Q_i^{ref}&=c Q_i^{pf} \\
V_{pp}^{ref}&=C_v C_q^{-1} Q_i^{ref}
\end{cases}
\end{equation} Now, one can compute the unknown terms, $Q_i^{ref}$ and $c$ in  \eqref{f2} by solving $\Sigma$ since it always provides the unique solution as there are $n+1$ equations and $n+1$ unknown variables in it. 
\subsection{Proposed SVC Strategy}
\begin{figure}[!htb]
	\centering
	\includegraphics[width=8cm]{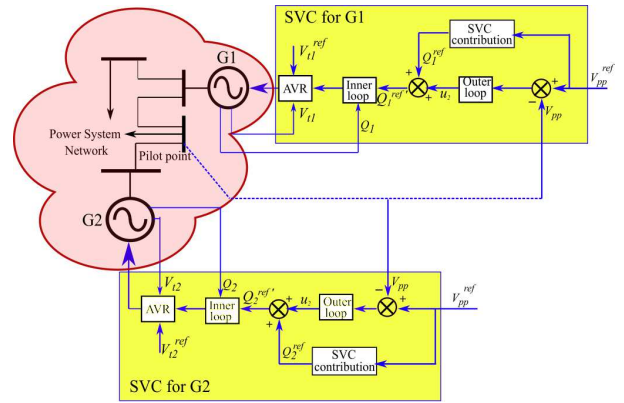}
	\caption{Overall block diagram of the proposed SVC strategy}
	\label{overall}
\end{figure}
The overall block diagram of the proposed SVC strategy is shown in Fig. \ref{overall}. In order to achieve the SVC objectives, the proposed SVC method is divided into two major control loops for its successful implementation. They are 
\begin{enumerate}
\item Inner loop : Reactive power control 
\item Outer loop : pilot point voltage control 
\end{enumerate}
Both inner and outer loop are time-decoupled to avoid unnecessary interaction between the loops. Inner loop is made faster than the outer loop in order to achieve the SVC objective. 
\subsubsection{Inner Loop}
\begin{figure}[!htb]
	\centering
	\includegraphics[width=8cm]{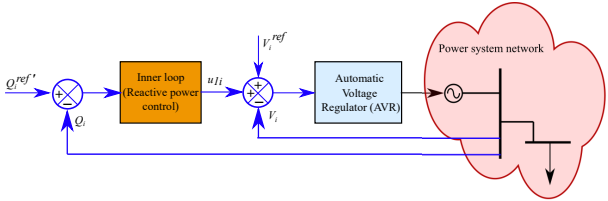}
	\caption{Block diagram of the inner loop: Reactive power control loop}
	\label{inner}
\end{figure}

This section mainly focuses on the design of the inner loop to achieve one of the primary objectives of SVC, i.e., reactive power alignment and control of each generators that contribute in the SVC action. The block diagram of the inner loop is shown in Fig. \ref{inner}. 
The inner loop mainly consists of the actual dynamics of power system where SVC action is carried out, AVR's which act as a primary voltage controller for regulating the terminal voltage $(V_{t_i})$ of each generator  and a reactive power controller. It is noted that primary AVR loop is also time-decoupled with the SVC inner loop. 
The reactive power reference signal  $Q_i^{ref}$ of each generator for the inner loop is generated by the outer loop and reactive power alignment block.The inner loop regulates the reactive power of the individual generators such that $V_{pp}(t)$ is maintained at $V_{pp}^{ref}$. 
\subsubsection{Outer Loop}
\begin{figure}[!htb]
	\centering
	\includegraphics[width=8cm]{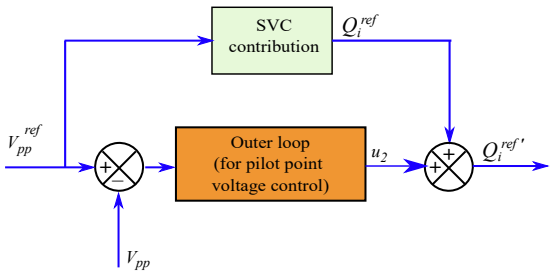}
	\caption{Block diagram of outer loop: Pilot point voltage control loop}
	\label{outer}
\end{figure}

This section mainly focus on the design of outer loop in order to achieve the main objective of SVC, i.e., set-point tracking  of pilot point voltage which is defined for a single zone. The block diagram of the outer loop along with reactive power alignment block is shown in Fig. \ref{outer}.
The outer loop mainly consists of pilot point voltage controller which generates a control signal, based on the information of $V_{pp}(t)$ and $V_{pp}^{ref}$. Further, this control signal from outer loop is added to the  output of reactive power alignment block to generate a reactive power reference signal $Q_i^{ref}$ for the individual generators. The generated reference signal $Q_i^{ref}$ is the input for the inner loop. This is shown in Fig. \ref{inner} and \eqref{f3}. 
There is a time-decoupling between the inner loop and the outer loop. Inner loop is operated faster (around 5 to 10 times) than the outer SVC loop. This is done because exact tracking of pilot point voltage is not possible unless the reactive power tracking in each generator reaches the steady state  according to the SVC objectives. 
\subsection{Controller Synthesis}
 DTiP controllers are designed and implemented for achieving the SVC objectives in each loop. DTiP1 controller is designed for reactive power control of each generators in inner loop. Similarly, for outer loop DTiP2 controller is designed for pilot point voltage control. Although different types of controller design and implementation approach can be used for this work,  a model-free control approach \cite{b8} is utilised for the implementation of the DTiP1 and DTiP2 controllers in this work. In model-free control, an ultra-local model is used instead of a complex differential equation that usually represents a complex system and the parameters of the model is continuously updated from the input and output measurements. This approach eliminates the requirement of a detailed mathematical model of the system and with a properly designed controller, it can perform well during internal parameter changes or external disturbances. Due to these features, model-free approach is adopted in this work.    For a single-input single-output system, the ultra-local model can be written as:
 \begin{equation}\label{m0}
	\dot{y}=F+\alpha u 
\end{equation} 
where $u$ and $y$ are the input and output variables respectively, and $\alpha \in \mathbb{R}$ is the design parameter which is selected by the designer is such a way that the magnitude of $\alpha$ is equal to the magnitude of $\dot{y}$. $F$ includes the information of any types of model uncertainties and disturbances acting on the system and $F$ is estimated from the $u$ and $y$ measurements.
 The controller is synthesised as an intelligent proportional (iP) or intelligent Proportional Integral (iPI) controller because of its simplicity in gain tuning compared to classic P and PI controllers. 
\begin{remark}
	\par Model-free control scheme was proposed for the first time in \cite{b8} and it has already been used by different authors and frequently implemented for power electronic applications \cite{b9,b16} and for regulating complex networks such as HVDC \cite{b4}  due to its inherent features such as great robustness against any types of uncertainties and external disturbances, simplicity and its effectiveness. 
\end{remark}
The brief discussion on the design procedure for the DTiP1 and DTiP2 controllers based on model-free control scheme is presented below.
\subsubsection{Design of DTiP1 controller for inner loop}
In case of inner loop, for tracking  of reactive power of each generator, the model-free control scheme is used since it requires only the information of output, i.e., $Q_i(k)$ and the control input $u_{1_i}(k)$ for its implementation. The control input $u_{1_i}(k)$ is generated with the help of DTiP1 controller. For inner reactive power control loop, the ultra-local discrete-time model is given by
\begin{equation}\label{m1}
	\dot{Q}_i(k)=F_{1_i}(k)+\alpha_{1_i}u_{1_i}(k),  ~~i=1, \cdots, n
\end{equation} where $\alpha_{1_i} \in \mathbb{R}$ is the design parameter which is selected by the designer is such a way that the magnitude of $\alpha_{1_i}u_{1_i}(k)$ is equal to the magnitude of $\dot{Q}_i(k)$. $F_{1_i}(k)$ includes the information of any types of model uncertainties and disturbances acting on the system. Based on model \eqref{m1}, the DTiP1 controller for reactive power controller in the inner loop is designed as
\begin{equation}\label{m2}
	u_{1_i}(k)=-\frac{F_{1_i}(k)-\dot{Q}_i^{ref'}(k)+K_{p_{1_i}}e_{1_i}(k)}{\alpha_{1_i}}
\end{equation} where $K_{p_{1_i}}$ is the tuning gain. $e_{1_i}(k)=Q_i(k)-Q_i^{ref'}$ is the error term which is set to be zero. \par For implementing the controller \eqref{m2}, one needs the information of unknown quantity $F_{1_i}(k)$ which is to be estimated. The estimated value of $F_{1_i}(k)$ is termed as $\bar{F}_{1_i}(k)$  which is computed as :
\begin{equation}
	\bar{F}_{1_i}(k)=\dot{Q}_i(k)-\alpha_{1_i}u_{1_i}(k-h_d)
\end{equation} where $u_{1_i}(k-h_d)$ is the control input delayed by the small amount $h_d$, which is a small lag time given to the controller $u_{1_i}(k)$ and is required for the estimation of $\bar{F}_{1_i}(k)$.
$\dot{Q}_i(k)$ is obtained with the help of numerical differentiation technique \cite{b19,b20}.

\begin{remark}
	Numerical differentiation technique is widely used for derivative estimation of noisy signals through an algebraic approach. It has received a lot of attention due to its remarkable features such as high precision, high accuracy in terms of reducing noise errors, and importance in the field of engineering like signal processing, image processing, automatic control and applied mathematics \cite{b21,b22}. It works well in noisy environment. The same algebraic differentiation technique is adopted in this work to compute the derivative of the required signals instead of normal derivative block available in the MATLAB Toolbox. For implementation of the numerical differentiation technique, one has to select the optimal values of two parameters, i.e., sampling time $(T_{ndf})$ and number of samples $(N_{ndf})$ in such a way that the length of the window time of the differentiators is too small. 
\end{remark}
\subsubsection{Design of DTiP2 controller for outer loop}
In case of the outer loop, for the tracking  of pilot point voltage, again model-free control scheme is used since it requires only the information of output, i.e., $V_{pp}(k)$ and the control input $u_2(k)$ for its implementation. The control input $u_2(k)$ is generated with the help of DTiP2 controller. The ultra-local discrete-time model for outer loop is given by
\begin{equation}\label{o1}
	\dot{V}_{pp}(k)=F_2(k)+\alpha_2u_2(k),  
\end{equation} where $\alpha_2 \in \mathbb{R}$ is the design parameter which is chosen by the designer is such a way that $\left|\alpha_2u_2(k)\right|= \left|\dot{V}_{pp}(k)\right|$. $F_2(k)$ consists of knowledge of any types of model uncertainties and disturbances acting on the system. Based on  \eqref{o1}, the DTiP2 controller for outer loop is designed as
\begin{equation}\label{o2}
	u_2(k)=-\frac{F_2(k)-\dot{V}_{pp}^{ref}(k)+K_{p_2}e_{2}(k)}{\alpha_{2}}
\end{equation} where $K_{p_{2}}$ is the tuning gain. $e_{2}(k)=V_{pp}(k)-V_{pp}^{ref}$ is the tracking error which is set to be zero. \par For implementing the controller \eqref{o2}, one needs the information of the unknown quantity, $F_{2}(k)$ which is to be estimated in the same manner as before. The estimated value of $F_{2}(k)$ is defined as $\bar{F}_{2}(k)$  which is calculated as :
\begin{equation}
	\bar{F}_{2}(k)=\dot{V}_{pp}(k)-\alpha_{2}u_{2}(k-h_d)
\end{equation} where $u_{2}(k-h_d)$ is the control input delayed by the small amount $h_d$, which is the small time lag given to the controller $u_{2}(k)$ for the computation of $\bar{F}_{2}(k)$.
$\dot{V}_{pp}(k)$ is obtained with the help of numerical differentiation technique as discussed earlier.
\par With the help of signal $Q_i^{ref}$ generated by the reactive power alignment block and control signal $u_2(k)$ from outer loop, finally, one can compute the reactive power reference signal $Q_{i}^{ref'}$ of each generator unit for inner loop  to achieve the set-point tracking of pilot point voltage along with the alignment of reactive power that contribute in the SVC action. It is expressed as :
\begin{equation} \label{f3}
Q_{i}^{ref'}=Q_i^{ref} + u_2
\end{equation}
By the use of \eqref{m2}, \eqref{o2} and \eqref{f3}, the SVC objectives can be easily achieved. The effectiveness of the proposed SVC strategy is demonstrated with the help of a benchmark system in the upcoming section.
\section{Simulation results}
\begin{figure}[!htb]
	\centering
	\includegraphics[width=8cm]{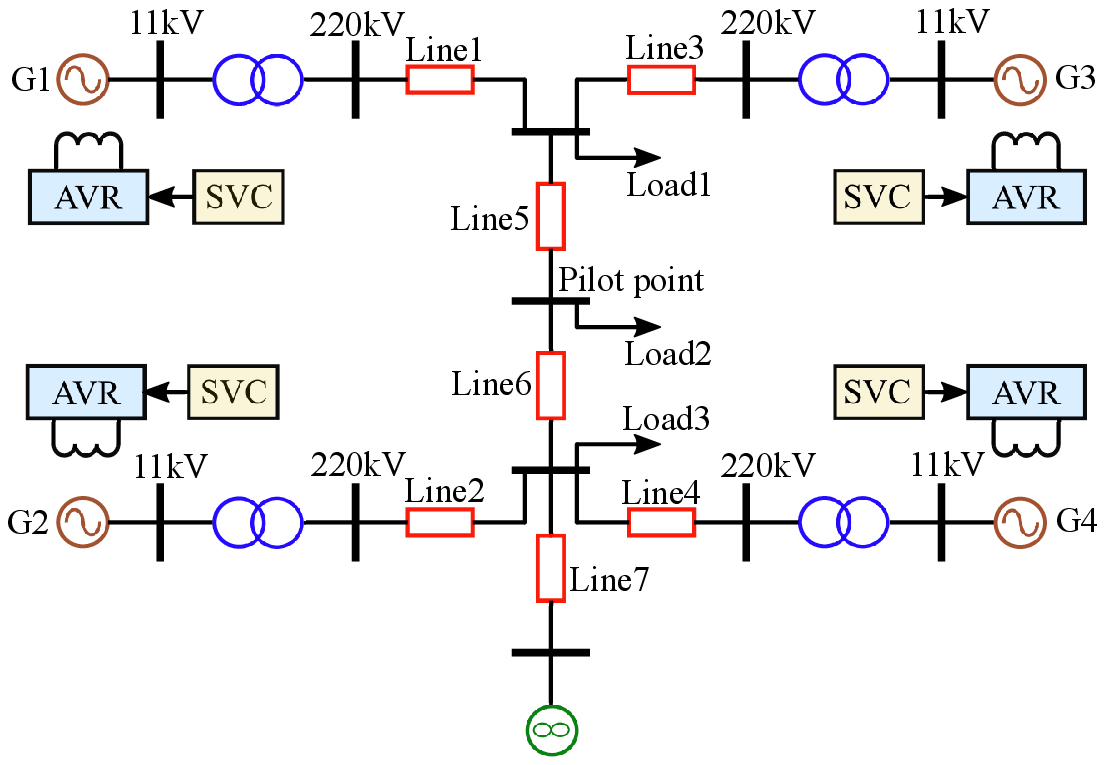}
	\caption{Schematic of the benchmark}
	\label{fig:fig3testsystem4gen}
\end{figure}

This section presents the description of a benchmark system used to validate  and show the efficacy of the newly proposed SVC scheme in this work.  The benchmark consists of a single zone of the power system with four synchronous generators (G1, G2, G3 and G4) of 200 MVA capacity each as shown in Fig. \ref{fig:fig3testsystem4gen}. The generators produce power at 11 kV, 50 Hz and a step up transformer increases the voltage  to 220 kV. The excitation system of each synchronous generator is controlled using an AVR.  The SVC attempts to provide correction signals to the AVRs to maintain the pilot point voltage of the zone at the reference value. There are three loads (Load1, Load2 and Load3) connected to the 220 kV line.  The benchmark system is modelled in MATLAB/Simulink \emph{Simscape} using discrete phasor simulation. The sample time for the power system elements, $T_{power}=10 ~\mu \text{s}$ and the sample time for the controller is  $T_{control}=100 ~\mu \text{s} $. 
\par The sensitivity matrices computed for the above benchmark system is given by:\\
$C_v=
\begin{bmatrix}
	0.2715 & 0.0989 & 0.2746 & 0.1022
\end{bmatrix}$,\\
$C_q=
\begin{bmatrix}
	+2.5370 & -0.3528 & -0.9798 & -0.3647\\
	-0.2729 & +2.8570 & -0.2761 & -0.6678\\
	-0.9774 & -0.3560 & +2.4910 & -0.3680\\
	-0.2729 & -0.6605 & -0.2823 & +2.7530
\end{bmatrix}$.
\par The corresponding values of tuned parameters of DTiP1 and DTiP2 controllers are given as :\\
$K_{p1}=
\begin{bmatrix}
	2 & 0 & 0 & 0\\
	0 & 2 & 0 & 0\\
	0 & 0 & 2 & 0\\
	0 & 0 & 0 & 2
	\end{bmatrix}$, $\alpha_1= \begin{bmatrix}
	4346 & 0 & 0 & 0\\
	0 & 4564 & 0 & 0\\
	0 & 0 & 4410 & 0\\
	0 & 0 & 0 & 4584
\end{bmatrix}$, $K_{p2}=0.09$, $\alpha_2=50000$. The following case studies are carried out on the benchmark system to demonstrate the performance of the proposed SVC strategy.

\subsection{Case 1: Step change in $V_{pp}^{ref}$: Results and discussion}
In this case, a step increase in $V_{pp}^{ref}$ is applied at 500 s, and the response of the pilot point voltage to this step change is observed with the proposed controller. 
\begin{figure}[!htb]
	\centering
	\includegraphics[width=6cm]{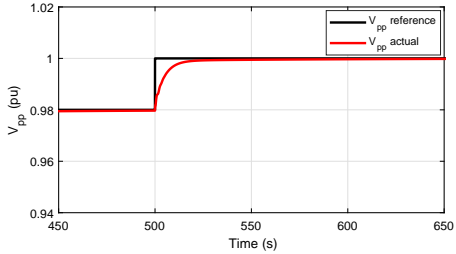}
	\caption{Pilot point voltage response for a step change  in $V_{pp}^{ref}$ }
	\label{fig:VpprefstepVppresponse}
\end{figure}
  \begin{figure}[!h]
	\centering
	\includegraphics[width=8cm]{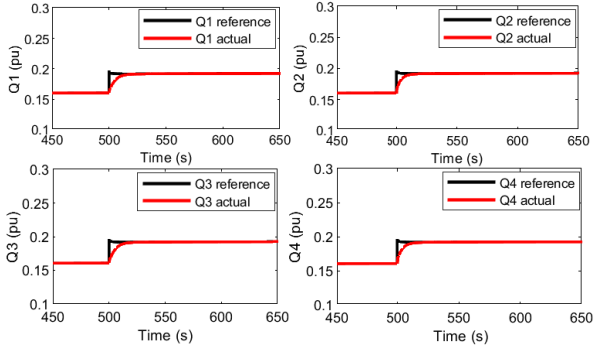}
	\caption{Reactive power response for a step change  in $V_{pp}^{ref}$ }
	\label{fig:VpprefstepQresponse}
\end{figure}

  \begin{figure}[!h]
	\centering
	\includegraphics[width=8cm]{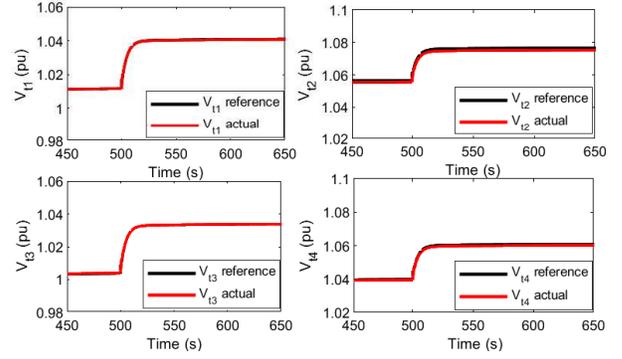}
	\caption{Terminal voltage response for a step change  in $V_{pp}^{ref}$ }
	\label{fig:VpprefstepVterminalresponse}
\end{figure}
Fig. \ref{fig:VpprefstepVppresponse} shows the response of the pilot point voltage during the step change in $V_{pp}^{ref}$. It can be observed from Fig. \ref{fig:VpprefstepVppresponse} that the actual pilot point voltage is tracking  $V_{pp}^{ref}$ during the step change.  Fig. \ref{fig:VpprefstepQresponse} shows the reactive power produced by the four generators during the step change in $V_{pp}^{ref}$. It can be seen from Fig. \ref{fig:VpprefstepQresponse} that the reference reactive power for the generators changes when $V_{pp}^{ref}$ changes and the actual reactive power is seen to track the reference reactive power. Fig. \ref{fig:VpprefstepVterminalresponse} shows the terminal voltage response of the four generators during the step change in $V_{pp}^{ref}$. The AVR control loop being faster than the SVC control loop, the terminal voltage tracks the reference signal quickly.
  \subsection{Case 2: Effect of delay in $V_{pp}$: Results and discussion}
\begin{figure}[!htb]
  	\centering
  	\includegraphics[width=6cm]{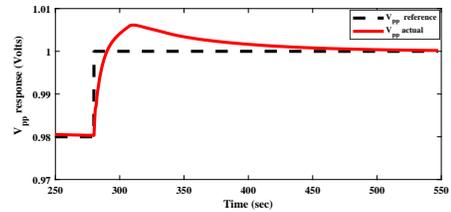}
  	\caption{Pilot point voltage response for a step change in $V_{pp}^{ref}$, with a delay of 28 s in $V_{pp}$}
  	\label{fig:stepchangeVpprefdel}
  \end{figure}

In this case, the behaviour of the benchmark system for a step change in $V_{pp}^{ref}$  with a delay in the measurement of $V_{pp}$ is analysed. The step change occurs at 280 s and the measurement delay of $V_{pp}$  is 28 s.  Fig. \ref{fig:stepchangeVpprefdel} shows the response of the pilot point voltage when $V_{pp}^{ref}$ changes from 0.98 pu to 1 pu at 280 s. It can be observed from Fig. \ref{fig:stepchangeVpprefdel} that $V_{pp}$ tracks $V_{pp}^{ref}$ is less than 200 s, even though there is a delay of 28 s in the measurement of $V_{pp}$. This indicates that the proposed control strategy is having reasonable resiliency against measurement delays in $V_{pp}$.
   
\begin{figure}[!htb]
  	\centering
  	\includegraphics[width=8cm]{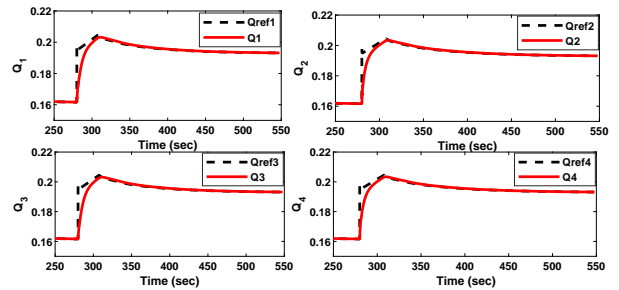}
  	\caption{Reactive power response for a step change  in $V_{pp}^{ref}$, with a delay of 28 s in $V_{pp}$}
  	\label{fig:VpprefwithdelayQresponse}
  \end{figure}

\begin{figure}[!htb]
  	\centering
  	\includegraphics[width=8cm]{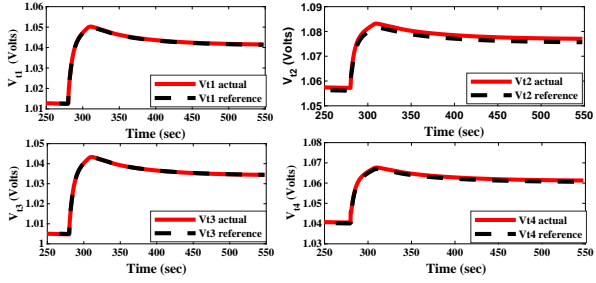}
  	\caption{Terminal voltage response for a step change in $V_{pp}^{ref}$, with a delay of 28 s in $V_{pp}$ }
  	\label{fig:VpprefwithdelayVresponse}
  \end{figure}

  Fig. \ref{fig:VpprefwithdelayQresponse} and Fig. \ref{fig:VpprefwithdelayVresponse} shows the reactive power  response and the terminal voltage response of the individual generators for this case.  The terminal voltages of the generators quickly track the reference values and this results in a good tracking performance of the reactive powers of the individual generators.
  
\subsection{Case 3: Load perturbation – Results and discussion}

\begin{figure}[!htb]
	\centering
	\includegraphics[width=6cm]{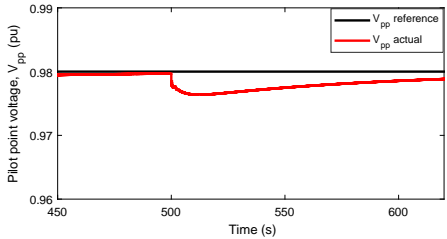}
	\caption{Pilot point voltage response for load perturbation }
	\label{fig:Vpprefloadperturbation}
\end{figure}

\begin{figure}[!htb]
	\centering
	\includegraphics[width=8cm]{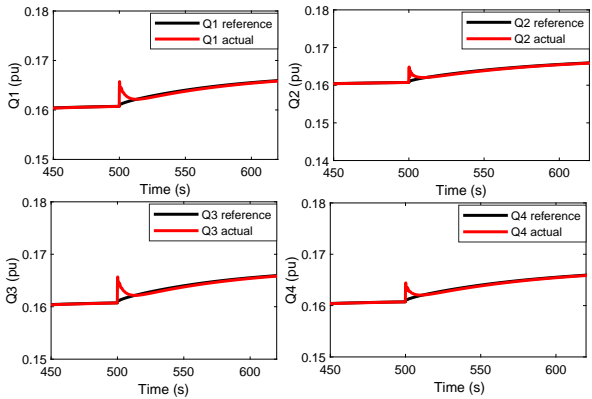}
	\caption{Reactive power response for load perturbation }
	\label{fig:Qloadperturbation}
\end{figure}

\begin{figure}[!htb]
	\centering
	\includegraphics[width=8cm]{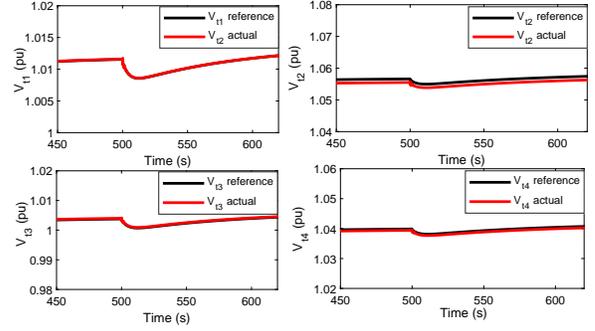}
	\caption{Terminal voltage response for load perturbation }
	\label{fig:Vtloadperturbation}
\end{figure}

In this case, the robustness of the controller for a step change in the load is analysed. At 500 s, an additional inductive load of 5 MVAR is connected at the pilot point and the response of the pilot point voltage to this change in load is observed. Fig. \ref{fig:Vpprefloadperturbation} shows the response of the pilot point voltage to this load perturbation. As seen in Fig. \ref{fig:Vpprefloadperturbation}, $V_{pp}$ drops from its steady state value after the load perturbation and later, returns to its previous value. The reactive power contributions of the individual generators are shown in Fig. \ref{fig:Qloadperturbation}. The reactive power references of the individual generators are seen to have changed after the load perturbation and the actual reactive power follows the reference signals quickly. The terminal voltage response of the generators are shown in Fig. \ref{fig:Vtloadperturbation}. The terminal voltages of the generators are seen to closely follow the terminal voltage references during the load perturbation.

\subsection{Case 4: Transmission line perturbation – Results and discussion}

In this case, the impact of change in transmission line parameters on the robustness of the proposed control strategy is analysed. At 500 s, the parameters of the transmission line connected to the generator 2 is perturbed by adding another transmission line of the same parameters, in parallel. At 650 s, the parallel connected transmission line is removed. The change in the transmission line parameters affects the voltage and reactive power contributions of the individual generators. 
\begin{figure}[!htb]
	\centering
	\includegraphics[width=6cm]{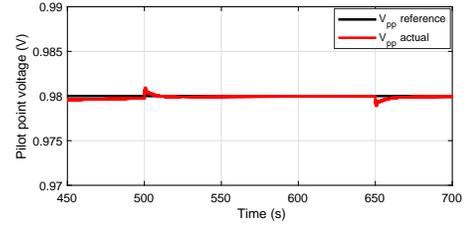}
	\caption{Pilot point voltage response for line perturbation }
	\label{fig:Vppreflineperturb}
\end{figure}
\begin{figure}[!htb]
	\centering
	\includegraphics[width=8cm]{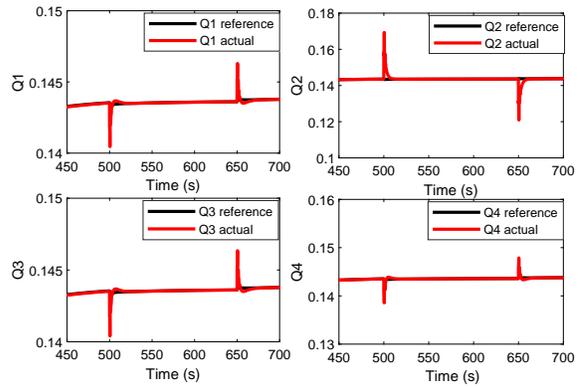}
	\caption{Reactive power response for line perturbation }
	\label{fig:Qresplineperturb}
\end{figure}
\begin{figure}[!htb]
	\centering
	\includegraphics[width=8cm]{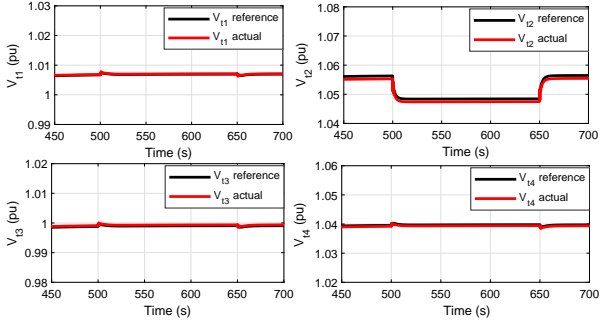}
	\caption{Terminal voltage response for line perturbation }
	\label{fig:Vtresplineperturb}
\end{figure}
Fig. \ref{fig:Vppreflineperturb} shows the response of the pilot point voltage during the line parameter perturbation. As seen from Fig. \ref{fig:Vppreflineperturb}, at the instant of parameter change at 500 s, there is an upward deviation in  $V_{pp}$. But $V_{pp}$ is seen to be returning to the earlier steady state value after the perturbation. At 650 s, there is a downward deviation in  $V_{pp}$ but it returns to the earlier steady state value after a transient period. Fig. \ref{fig:Qresplineperturb} and Fig. \ref{fig:Vtresplineperturb} show the reactive power response and terminal voltage response, respectively, of  the individual generators during the line parameter perturbation. The results indicate the robust performance of the proposed control strategy during disturbances.

\subsection{Case 5: Generator disconnection – Results and discussion}

In this case, the impact of disconnection of an individual generator participating in the secondary voltage control is studied.
\begin{figure}[!htb]
	\centering
	\includegraphics[width=6cm]{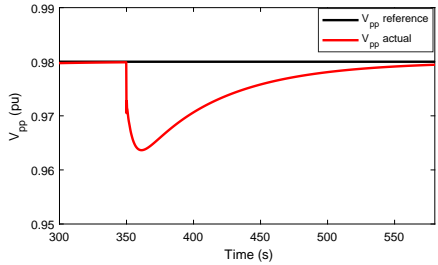}
	\caption{Pilot point voltage response for generator disconnection }
	\label{fig:Vpprefgendisconnection}
\end{figure}
\begin{figure}[!h]
	\centering
	\includegraphics[width=6cm]{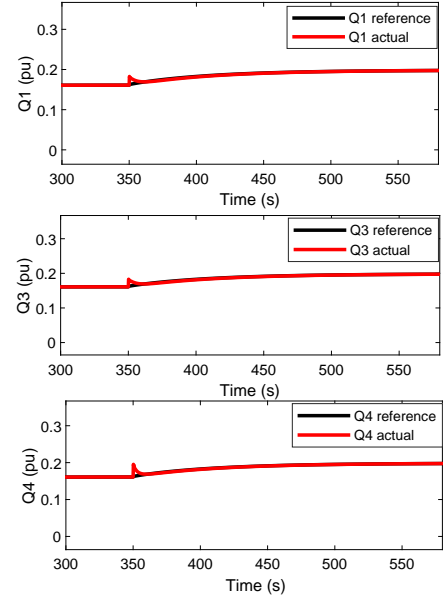}
	\caption{Reactive power response for generator disconnection }
	\label{fig:Qrespgendisconnection}
\end{figure}
\begin{figure}[!h]
	\centering
	\includegraphics[width=6cm]{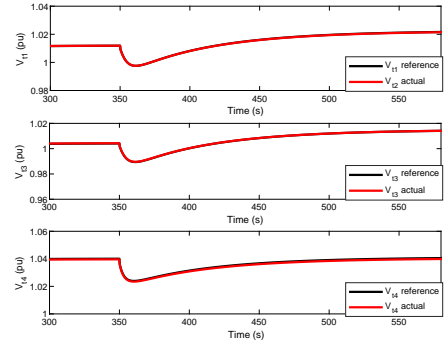}
	\caption{Terminal voltage response for generator disconnection }
	\label{fig:Vtrespgendisconnection}
\end{figure}
At 350 s, the generator G2 is disconnected and the response of the pilot point voltage is shown in Fig. \ref{fig:Vpprefgendisconnection}. When the generator G2 is disconnected, the pilot point drops momentarily. With the proposed SVC strategy, the pilot point voltage returns to the reference value in less than 250 s after the generator disconnection, as shown in Fig. \ref{fig:Vpprefgendisconnection}. The regulation of the pilot point voltage was possible with the remaining three generators supplying additional reactive power as seen in Fig. \ref{fig:Qrespgendisconnection}.
Fig. \ref{fig:Vtrespgendisconnection} shows the terminal voltage responses of the remaining three generators. After the generator G2 disconnection, the terminal voltages of the remaining three generators (G1, G3 and G4) reaches a new steady state value to compensate for the loss of generator G2.
  \subsection{Case 6: Delayed participation of a generator in SVC – Results and discussion}
In this case, the impact of one generator (G2)  participating in the SVC after a delay period is analysed. Initially, all the generators except G2 participates in SVC. At 350 s, there is a step increase in $V_{pp}^{ref}$ and at 500 s, G2 starts the participation in the SVC.

\begin{figure}[!htb]
	\centering
	\includegraphics[width=7cm]{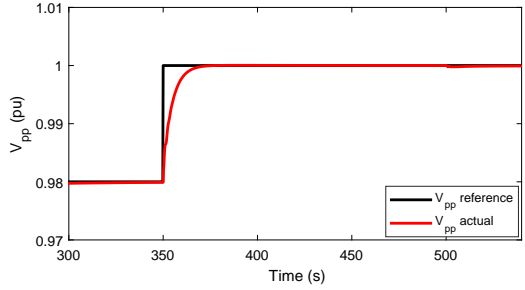}
	\caption{Pilot point voltage response for a delayed participation of a generator in SVC}
	\label{fig:Vppintegration SVC}
\end{figure}

\begin{figure}[!htb]
	\centering
	\includegraphics[width=8cm]{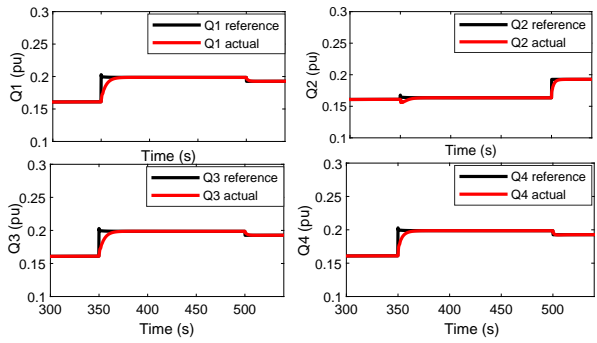}
	\caption{Reactive power response for a delayed participation of a generator in SVC integration to SVC }
	\label{fig:Qintegration SVC}
\end{figure}

\begin{figure}[!htb]
	\centering
	\includegraphics[width=8cm]{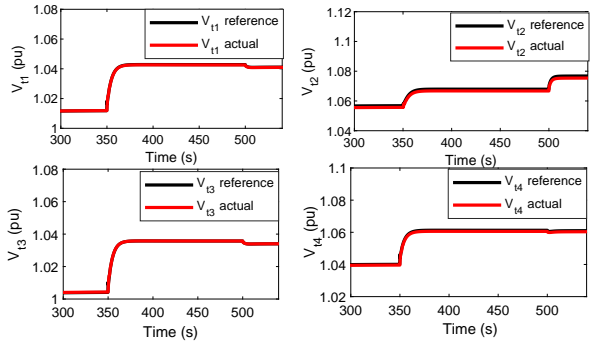}
	\caption{Terminal voltage response for a delayed participation of a generator in SVC}
	\label{fig:Vtintegration SVC}
\end{figure}

Fig. \ref{fig:Vppintegration SVC}  shows the response of the pilot point voltage. It can be seen that $V_{pp}$ follows $V_{pp}^{ref}$ after 350 s, when $V_{pp}^{ref}$ changes from 0.98 pu to 1 pu. When G2 starts participating in the SVC at 500 s, there is a minor dip in $V_{pp}$ which is quite negligible. Fig. \ref{fig:Qintegration SVC} shows the reactive power generations of the individual generators. Just after 350 s, G1, G3 and G4 starts contributing reactive power. At 500 s, when G2 starts its participation in SVC, the reactive power contributions of the individual generators are modified due to the contribution of G2. Once G2 starts contributing for SVC, the reactive power generations of G1, G3 and G4 reduces compared to their values before 350 s. Fig. \ref{fig:Vtintegration SVC} shows the terminal voltages of the individual generators for the delayed participation of G2 in SVC. After 350 s, the terminal voltages of the generators increases due to increase in the reference pilot point voltage. The rise in terminal voltage of the G2 is less after 350 s, when it is not participating in SVC. After 500 s, the terminal voltage of G2 rises further due to its particpation in SVC, while the terminal voltage of other generators reduces slightly. The results indicate that the proposed control strategy works satisfactorily when a non-participating generator in the SVC starts its participation at a later stage. 

\section{Conclusions}
The decentralised SVC control scheme proposed in this paper eliminates the limitations of the classic centralized SVC control schemes related to robustness or the use of advanced controls which relay on detailed models of the system which need to be reconfigured. \\

Also, the new inner-outer loops structure proposed here allows one to have independent controls for each generator. This increases resilience especially in case of renewable generators which may often come out and in the control, in function of evolution of the natural resources (wind and sun). Such plug and play facility is a major advantage for integration of renewables to secondary voltage ancillary services. Indeed, our control allows such generators to participate in classic secondary voltage controls on equal position with the classic thermal generators. This topic is developed in POSYTYF H2020 project by considering the renewables grouped into the new concept of Dynamic Virtual Power Plant \cite{DVPP_WPLs}. \\

This scheme has proved increased robustness against load and grid variations as well as against transmission delays, which are classic requirements for voltage secondary controls. \\

Although the proposed control scheme is  implemented in this work using model-free control, it can be implemented with different controllers also. \\

The future scope lies on extension of the proposed control scheme to larger grids with multiple SVC zones and larger number of renewable energy sources participating in the SVC. Also, hardware in the loop validation is scheduled.

\section*{Acknowledgment}
This project has received funding from the European Union's Horizon 2020 research and innovation programme under grant agreement No. 883985 (POSYTYF - POwering SYstem flexibiliTY in the Future through RES, https://posytyf-h2020.eu/)



\end{document}